\documentclass{aastex}
\usepackage{spr-astr-addons}
\usepackage{url}

\RequirePackage{color}

\newcommand{\emaila}{krisztina.g@gmail.com}

\begin{document}

\title{The rare extended radio-loud narrow-line Seyfert 1 galaxy SDSS J1030+5516 at high resolution}
\slugcomment{}
\shorttitle{SDSS J1030+5516 at high resolution}
\shortauthors{Gab\'anyi et al.}

\author{K. \'E. Gab\'anyi\altaffilmark{1}}
\email{\emaila}
\affil{MTA-ELTE Extragalactic Astrophysics Research Group, \emaila}
\and 
\author{S. Frey} 
\affil{Konkoly Observatory, MTA Research Centre for Astronomy and Earth Sciences}
\and 
\author{P. Veres} 
\affil{Center for Space Plasma and Aeronomic Research, University of Alabama in Huntsville}
\and
\author{A. Mo\'or}
\affil{Konkoly Observatory, MTA Research Centre for Astronomy and Earth Sciences}

\altaffiltext{1}{Konkoly Observatory, MTA Research Centre for Astronomy and Earth Sciences}

\begin{abstract}
Recently, \citet{disc} reported the discovery of SDSS J103024.95$+$551622.7, a radio-loud narrow-line Seyfert 1 galaxy having a $\sim 100$\,kpc scale double-lobed radio structure. Here we analyse archival radio interferometric imaging data taken with the Very Large Array (VLA) at 5~GHz, and with the Very Long Baseline Array (VLBA) at 4.3 and 7.6~GHz. Two hotspots and a compact core are detected with the VLA at arcsec scale, while a single milliarcsec-scale compact radio core is seen with the highest resolution VLBA observations. The {\em Fermi} Large Area Telescope did not detect $\gamma$-ray emission at the position of this source. In the mid-infrared, the {\em Wide-field Infrared Survey Explorer} satellite light curve, covering more than 7 years and including the most recent data points, hints on flux density variability at 3.4\,$\mu$m. Our findings support the notion that this source is a young version of Fanaroff--Riley type II radio galaxies.
\end{abstract}



\section{Introduction}

Narrow-line Seyfert 1 galaxies (NLS1) form a peculiar subclass of active galactic nuclei (AGN). They are defined by their narrow permitted optical lines \citep[the full width at half maximum, FWHM, of H$\beta$ line is below $2000$\,km\,s$^{-1}$,][]{goodrich1989}, a flux ratio of [\ion{O}{3}]$\lambda5007$ to H$\beta$ smaller than 3 \citep{OsterbrockPogge1985}, and the strong emission of the \ion{Fe}{2} multiplets. However, \cite{Cracco2016} showed that having strong iron lines may not be a distinctive property of NLS1 sources in accordance with the study of quasar emission lines of \cite{BorosonGreen}, which showed the anticorrelation between the strength of \ion{Fe}{2} and \ion{O}{3} lines.

The narrow permitted lines of NLS1 sources are explained with their relatively lower-mass central black holes, $10^6-10^8$M$_\odot$ \citep{Mathur2000}, which consequently means high accretion rates close to the Eddington limit \citep{BorosonGreen}. Based upon these, \cite{Mathur2000} proposed that NLS1 sources can be young AGN residing in rejuvenated galaxies. Alternatively, the narrow H$\beta$ lines of NLS1 sources can be due to orientation effect, if their disk-like broad-line regions are seen pole-on \citep{Decarli2008}.

Similarly to AGN in general, a small fraction, $\sim 7$\% of NLS1 sources are radio-loud \citep{Komossa2006,Zhou2006}, where radio-loudness is determined by the ratio of the $6$\,cm radio to the $4400$\,\AA \,optical flux density following \citet{radioloudness}. \citet{VeereshSingh2018} studied the radio properties of a large sample of optically-selected NLS1 sources, and found that the 
radio-detected ones have small sizes, $<30$\,kpc. The most radio-loud NLS1 sources \citep[and the ones with the highest radio luminosity investigated by][]{VeereshSingh2018} show blazar-like properties: flat radio spectrum, compact radio cores, high brightness temperatures, significant variability, flat X-ray spectra, and altogether blazar-like spectral energy distribution \citep[SED; e.g.][]{Yuan2008}. 
Several of them were also detected in $\gamma$-rays with the {\em Fermi} satellite \citep[for a full list, see][]{Romano2018}, 
and in a few of them superluminally moving radio jet components were imaged with very long baseline interferometry (VLBI) technique \citep{Lister2016}. Therefore these sources, similarly to blazars, are thought to possess relativistic radio jets inclined at small angle to the line of sight. 

Few of the radio-loud NLS1 sources have kpc-scale (from a few tens of kpc to $\sim 100$\,kpc) radio structures. \citet{doi-kpc} found that the detection rate of extended radio emission in NLS1 sources is lower than in broad-line AGN. 
This is confirmed more recently by \cite{Berton2018}. They observed $74$ NLS1 sources and found that the majority of flat-spectrum radio-loud NLS1 sources have compact morphology on kpc scale. In most of the extended radio-loud NLS1 sources, the radio emission is two-sided \citep{doi-kpc,Richards_kpc,Congiu-kpc,J1100}.

Recently, \citet{disc} reported the discovery of a radio-loud NLS1 source, \\ SDSS J103024.95$+$551622.7 (hereafter J1030$+$5516) with arcsec-scale structure similar to those of Fanaroff--Riley type II radio galaxies \citep[FR\,II,][]{fr}. The projected linear size, $\sim 110$\,kpc is among the largest values in radio-loud NLS1 sources. Using low-resolution ($\sim 5\arcsec$) radio data of J1030$+$5516 from the Faint Images of the Radio Sky at Twenty-Centimeters (FIRST) survey \citep{FIRST}, \citet{disc} argue that the inclination angle of the jet in the core region is $<12\degr$ with respect to the line of sight. 

Here we present sub-arcsec resolution archival Very Large Array (VLA) A-configuration data, and milli-arcsec (mas) resolution VLBI data of J1030$+$5516, which support the claims of \citet{disc}. Additionally, we analysed more than $10$\,yr of {\it Fermi} Large Area Telescope \citep[LAT; ][and references therein]{Fermi_4cat} data to constrain the high-energy properties of the source. We also re-evaluated the mid-infrared light curves covering more than 7\,yr, obtained with the {\em Wide-field Infrared Survey Explorer} \citep[{\em WISE},][]{wise} satellite.

In the following, we assume a flat $\Lambda$CDM cosmological model with $H_0=70$\,km\,s$^{-1}$\,Mpc$^{-1}$ and $\Omega_\textrm{m}=0.27$. At the redshift of J1030$+$5516, $z=0.435$, $1\arcsec$ angular size corresponds to a projected linear size of $5.65$\,kpc.

\section{Observing data}

\subsection{Archival VLA radio data}

J1030$+$5516 was observed at $5$\,GHz with the VLA in its most extended A configuration on 1992 October 20 (project code: AF233). The raw data were obtained from the US National Radio Astronomy Observatory (NRAO) archive\footnote{\url{https://archive.nrao.edu/}}. The on-source integration time was 1\,min, the total bandwidth was 100\,MHz. Phases and amplitudes were calibrated in the NRAO Astronomical Image Processing System \citep[{\sc AIPS},][]{aips} in a standard way. The flux density scale was set using the amplitude calibrator source 3C286. We used the {\sc Difmap} \citep{difmap} software for imaging and for fitting Gaussian brightness distribution model components directly to the interferometer visibility data.

\subsection{Archival VLBA radio data}

J1030$+$5516 was observed with the Very Long Baseline Array (VLBA) on 2016 May 28 at $4.3$ and $7.6$\,GHz (project code: BP192, PI: L. Petrov) in the framework of the wide-field VLBA calibrator survey (L. Petrov 2019, in preparation). Nine (Brewster, Fort Davis, Hancock, Kitt Peak, Los Alamos, North Liberty, Owens Valley, Pie Town, St. Croix) and eight (all the above but Hancock) antennas of the array were used at the lower and higher frequency, respectively. The bandwidth was 256\,MHz and the integration time was nearly 1\,min at both frequencies. The calibrated visibilities were obtained from the Astrogeo website\footnote{\url{http://astrogeo.org/vlbi_images}, maintained by L. Petrov}.  

\subsection{Archival {\em Fermi}/LAT $\gamma$-ray data}

We analysed the archival {\em Fermi}/LAT data of J1030$+$5516. We looked for $\gamma$-ray signals in the available $10.3$-year data which cover the time range between 2008 August 4 and 2018 November 26. We derived {\it Fermi}/LAT upper limits using the routines included in the {\tt fermipy} package \citep{Wood+17fermipy}.  We selected a 15$^{\circ}$ circular region around the position of J1030+5516, and  an energy range of $0.1-100$\,GeV. We used Pass 8, SOURCE type photons, with {\it P8R2\_SOURCE\_V6} responses. The fit included a diffuse galactic foreground (gll\_iem\_v6) and an isotropic component (iso\_P8R2\_SOURCE\_V6\_v06).

\subsection{{\it WISE} data}

The mid-infrared {\it WISE} satellite scanned the whole sky in four bands at $3.4$, $4.6$, $12$, and $22\mu$m (referred as W1, W2, W3, and W4) during its original mission phase in 2010 \citep{wise}. Afterwards, the satellite measurements are continued within the framework of the NEOWISE (Near-Earth Object WISE) project \citep{neowise}. After four months of NEOWISE observations, the satellite was hibernated for 34 months. Then the NEOWISE Reactivation Mission continued. In this currently on-going phase, observations are conducted only at the two shorter wavelength bands, since the cooling material required for W3 and W4 receivers has been depleted. The {\it WISE} satellite observes the same regions of the sky in every $\sim 180$\,days. 

We downloaded the {\it WISE} single exposure data\footnote{\url{http://irsa.ipac.caltech.edu/Missions/wise.html}} up until 2017 November, and followed the procedure as in \cite{wise_nls1}. We adapted the guidelines in the Explanatory Supplement Series\footnote{\url{http://wise2.ipac.caltech.edu/docs/release/allwise/expsup/sec3_2.html}} to discard bad quality data points. None of the measurements were affected by the South-Atlantic Anomaly, or scattered light from the Moon. We used only those measurements for which the frame image quality score (`{\verb!qi_fact!}') was $1.0$, since values less than $1.0$ mark data where residual light system motion may degrade the flux measurements. Ten per cent of data were discarded because of this effect. The contamination and confusion flags (`{\verb!cc_flags!}') did not indicate problems for any of the measurements.

The final light curve contains $177$ points grouped into ten mission phases both in W1 and W2 bands. Each mission phase lasted usually for $\lesssim2$\,days and contains $12-16$ points, except for one mission phase, which is the combination of two $\sim1$-day long observations separated by $2$ days and contains $26$ data points.

\section{Results}

\subsection{VLA data}

\begin{figure}[t]
\includegraphics[width=\columnwidth, bb=0 60 510 510, clip=]{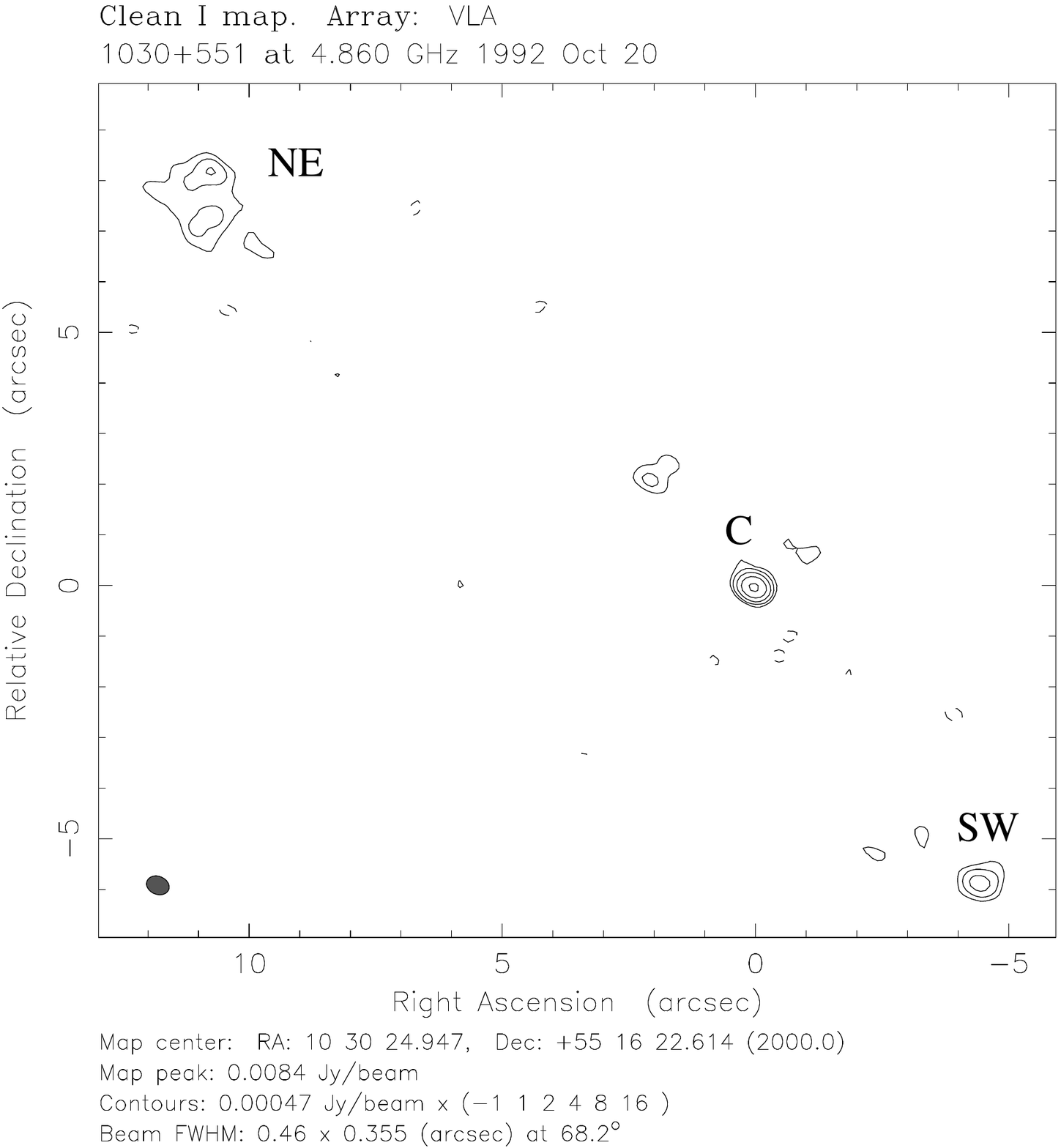}
\caption{%
5-GHz VLA image of J1030$+$5516 taken on 1992 October 20. The image is centred at the brightest pixel at right ascension $10^\mathrm{h} 30^\mathrm{m} 24\fs947$ and declination  $55\degr 16\arcmin 22\farcs6$. Peak brightness is $8.4$\,mJy\,beam$^{-1}$, the lowest contours are drawn at $\pm0.5$\,mJy\,beam$^{-1}$, corresponding to $\sim 3\sigma$ image noise level. (Dashed lines represent negative contours.) The positive contours increase by a factor of 2. The restoring beam is $0\farcs5\times0\farcs4$ (FWHM) with a major axis position angle $68\degr$, as shown in the lower left corner of the image. The three fitted components listed in Table \ref{tab:VLA_model} are labeled} 
\label{fig:VLA}
\end{figure}

\begin{table*}
\small
\caption{Parameters of the circular Gaussian components fitted to the $5$-GHz VLA visibilities\label{tab:VLA_model}}
\begin{center}
\begin{tabular}{@{}ccccc}
\tableline
ID & Flux density & Relative RA & Relative Dec. & FWHM size \\
 & (mJy) & (arcsec) & (arcsec) & (mas) \\
\tableline
C & $9.3 \pm 0.4$ & -- & -- & $86\pm37$ \\ 
SW & $4.8\pm 0.6$ & $-4.47 \pm 0.03$ & $-5.84\pm0.03$ & $346\pm52$ \\
NE & $13.7\pm1.7$ & $10.87 \pm 0.08$ & $7.67\pm0.07$ & $1386\pm140$ \\
\tableline
\end{tabular}
\end{center}
Notes: component name in Col.~1, flux density in Col.~2, offset in right ascension and declination with respect to the central component C in Cols.~3 and 4, and the FWHM size in Col.~5
\end{table*}

The $5$-GHz VLA-A observation revealed three distinct features, the core, the southwest (SW) and the northeast (NE) lobes (Fig.~\ref{fig:VLA}). This structure is in good agreement with the lower-resolution FIRST image presented by \citet{disc}. Three circular Gaussian model components can adequately describe the radio structure. Their parameters are given in Table~\ref{tab:VLA_model}. The distance between the components fitted to the hotspots in the SW and NE lobes is $20\farcs43 \pm 0\farcs01$, which agrees well with the $20\farcs5$ source size at $1.4$\,GHz, as derived by \citet{disc} from the FIRST image.

Using the overall spectral index value $\alpha = -0.65$ (defined as $S\propto\nu^\alpha$, where $S$ is the flux density and $\nu$ the frequency) and the integrated flux density $S_{1.4} = 155$\,mJy derived from the $1.4$-GHz FIRST data by \citet{disc}, the flux density at $5$\,GHz can be expected as $\sim68$\,mJy. The sum of flux densities of the fitted components (Table~\ref{tab:VLA_model}), $S_5 = 27.8$\,mJy, is well below this value, indicating that a significant amount of diffuse radio emission was resolved out in the 5-GHz VLA-A observation.

\subsection{VLBA data}

At both frequencies, a single unresolved component was detected with the VLBA. There was no additional radio-emitting feature down to $0.8$\,mJy\,beam$^{-1}$ within the undistorted field of view with a radius of $0\farcs3$ at $4.3$\,GHz, and down to $1.1$\,mJy\,beam$^{-1}$ within the undistorted field of view with a radius of $0\farcs2$ at $7.6$\,GHz.

We used the {\sc Difmap} \citep{difmap} software to fit the visibilities with a circular Gaussian brightness distribution. At $4.3$\,GHz, the radio emission can be best described by a point source model. During model-fitting using a Gaussian brightness distribution as a starting model, the FWHM size of the feature converged to an unrealistically small value ($\sim 10^{-6}$\,mas), indicating the unresolved nature of the detected component. The flux density of the point source model component is $S_{4.3} = 10.6 \pm 0.3$\,mJy. 

At $7.6$\,GHz, a stable fit could be reached with using a single circular Gaussian component. Its parameters are: flux density $S_{7.6}=15.2 \pm 0.5$\,mJy, FWHM diameter $\theta=0.31 \pm 0.09$\,mas. However, this size is still smaller than the minimum resolvable angular size of the interferometer array, $0.4$\,mas, calculated following the formula of \citet{kovalev_size}. Therefore J1030$+$5516 remained unresolved with VLBI at $7.6$\,GHz as well. 

Using the $0.4$\,mas upper limit to the source size, we can calculate a lower limit to the brightness temperature:
\begin{equation}
T_\mathrm{B}=1.22 \times 10^{12} (1+z) \frac{S}{\theta^2\nu^2} \mathrm{K},
\end{equation}
where $z$ is the redshift, $S$ is the flux density in Jy, $\nu$ is the observing frequency in GHz, and $\theta$ is the FWHM size of the Gaussian component in mas. Thus, the lower limit to the brightness temperature of J1030$+$5516 is $T_\mathrm{B} \geq 3\times 10^9$\,K. This value is below the equipartition brightness temperature limit \citep[$\sim 5 \times 10^{10}$\,K;][]{readhead} by an order of magnitude. However, since it is a lower limit only, it does not exclude the possibility of Doppler boosting caused by relativistic beaming in the jet. More sensitive and higher-resolution VLBI data would be needed to place tighter constraints on the angular size of the compact central source in J1030$+$5516 and thus on the brightness temperature. 

Using these simultaneous dual-frequency VLBA observations, we can derive the two-point spectral index $\alpha_{4.3}^{7.6}=0.6\pm0.2$ of the core,  indicating an inverted radio spectrum of the most compact component. This is not in contrast with the spectral index derived by \citet{disc} ($\alpha = -0.65 \pm 0.04$), since that value was obtained using low-resolution radio observations. Those measure the total flux density of J1030$+$5516 which is dominated by the steep-spectrum lobes and diffuse emission, completely resolved out on the long baselines of the VLBA.

\subsection{{\em Fermi}/LAT data}

\begin{table}
\small
\caption{{\em Fermi} upper limits of the $\gamma$-ray flux of J1030$+$5516 \label{tab:fermi}}
\begin{center}
\begin{tabular}{@{}cc}
\tableline
Energy range & Flux \\
 (GeV) & (MeV\,cm$^{-2}$\,s$^{-1}$) \\
\tableline
$0.1 - 0.316$ & $<4.5 \cdot 10^{-7} $ \\
$0.316-1$ & $<3.9 \cdot 10^{-8} $ \\
$1-3.162$ & $<6.8 \cdot 10^{-8} $ \\ 
$3.162-10$ & $<2.8 \cdot 10^{-8} $ \\
$10-31.622$ & $<6.6 \cdot 10^{-8} $ \\
$31.622-100$ & $<1.4 \cdot 10^{-7} $ \\ 
\tableline
\end{tabular}
\end{center}
\end{table}

No $\gamma$-ray emitting source was detected in the available {\em Fermi}/LAT data at the position of J1030$+$5516. The upper limits of the $\gamma$-ray fluxes in six energy ranges are given in Table~\ref{tab:fermi}.

\subsection{{\it WISE} mid-infrared lightcurve}

\begin{figure}[t]
\includegraphics[width=\columnwidth, bb=35 25 730 510, clip=]{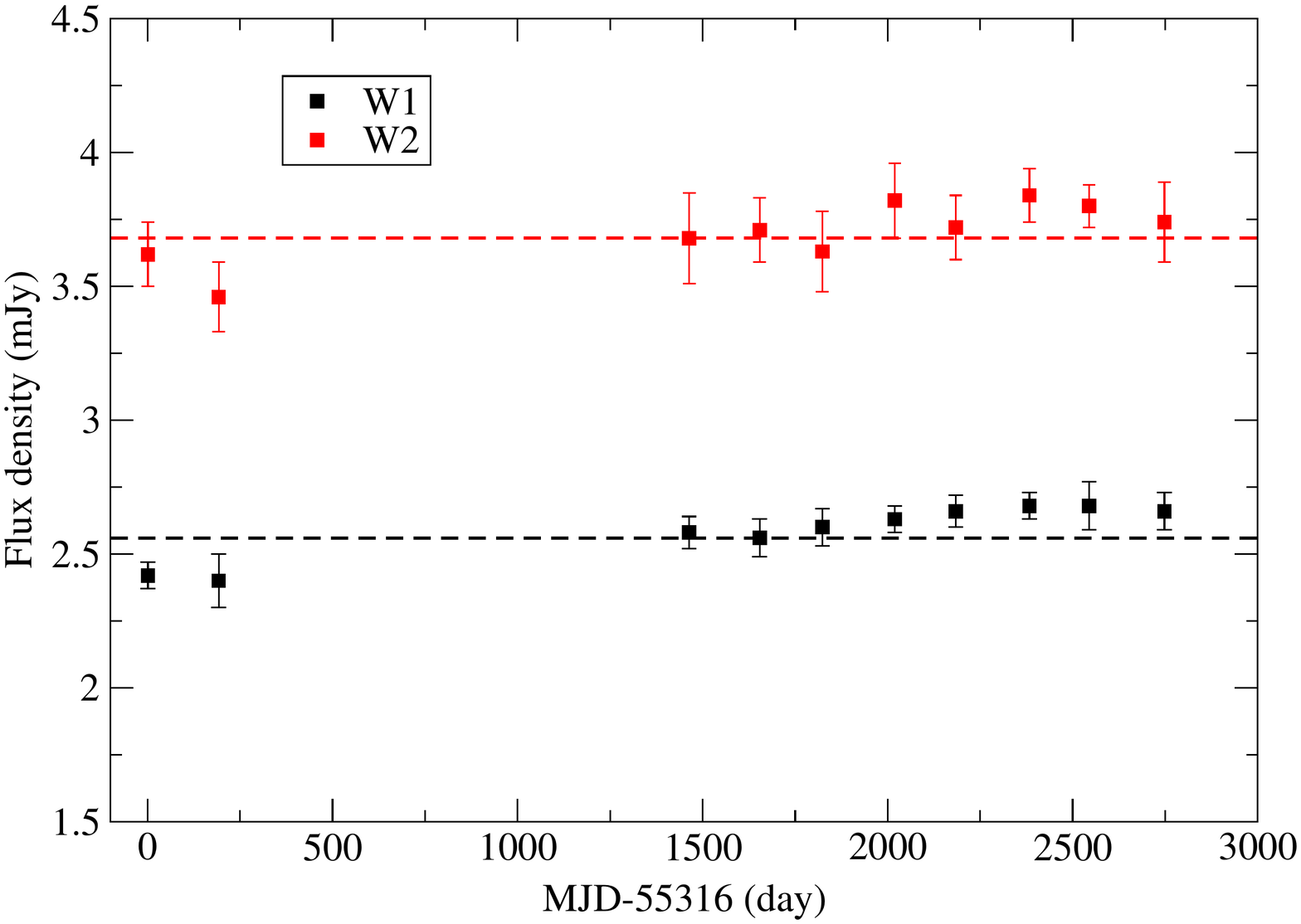}
\caption{%
Infrared light curves of J1030$+$5516, measured by the {\em WISE} satellite. Black and red symbols represent measurements made at $3.4$\,$\mu$m (W1) and $4.6$\,$\mu$m (W2), respectively. The symbols are average values of each mission phase, the error bars represent the variability within each mission phase. The dashed lines show the long-term average flux densities in the two bands} 
\label{fig:wise}
\end{figure}

We converted the {\it WISE} magnitudes to flux densities, following the description in the Explanatory Supplement Series.\footnote{\url{http://wise2.ipac.caltech.edu/docs/release/allsky/expsup/sec4_4h.html}} To investigate the variability, we calculated the reduced $\chi^2$ for all the data in a given band, and for each mission phase, separately. For every single mission phase, the reduced $\chi^2\lesssim2$, indicating no variability on a few-day long time scale. The reduced $\chi^2$ values for all measurements are $3.3$ at $3.4\mu$m and $2.8$ at $4.6\mu$m, showing a hint of variability at the shorter wavelength. 

\citet{disc} also analysed the infrared light curve of J1030$+$5516 measured with the {\it WISE} satellite. At the time of their publication only eight mission phases were available, until 2016 November. They found that the object is not variable at $3.4\mu$m and $4.6\mu$m. We also calculated the reduced $\chi^2$ using the first eight mission phases used by \cite{disc}. We found that the reduced $\chi^2$ are lower for both bands, $3.1$ at $3.4\mu$m and $2.4$ at $4.6\mu$m. 

In Fig. \ref{fig:wise}, we plot the weighted average flux densities and the standard deviations for each mission phase in both bands.

\section{Discussion}

The obtained lower limit of the brightness temperature of the mas-scale radio emitting core of J1030$+$5516 agrees well with the values derived by \citet{Gu2015} for a sample of $14$ radio-loud NLS1 sources, $10^{8.4}\,\mathrm{K} < T_\mathrm{B} < 10^{11.4}\,\mathrm{K}$. These low values compared to powerful blazar jets are explained by intrinsically low jet power by \citet{Gu2015}. The brightness temperature measured in J1030$+$5516 is also similar to another NLS1 source with a radio structure extended to $\sim 150$\,kpc, SDSS J110006.07$+$442144.3 \citep{J1100}. 
\citet{Richards_kpc} investigated three radio-loud NLS1 galaxies with kpc-scale radio structures and found mildly relativistic jets. Comparing the mas-scale structures, J1030$+$5516 is more compact and its brightness temperature is $3-20$ times larger than the three NLS1 sources studied by \citet{Richards_kpc}. 

The flux density of the core component measured at $5$\,GHz with the VLA at arcsec scale ($9.3\pm0.4$\,mJy) is below the value measured at mas-scale resolution with the VLBA at a slightly lower frequency of $4.3$\,GHz ($10.6\pm0.3$\,mJy). As this cannot be explained by resolution effect, it is more likely related to variability of the radio flux density. 

In the standard picture of expanding radio galaxies, the lobe located closer to the observer is seen farther away from the host galaxy in projection \citep{arm-length}, thus the arm-length ratio of the brighter to the fainter lobe is larger than one. In J1030$+$5516, the distance between the brighter NE lobe and the central compact radio feature, C is $d_\mathrm{NE}=13\farcs3\pm0\farcs1$, while the fainter SW lobe is at $d_\mathrm{SW}=7\farcs35\pm0\farcs04$. Therefore, it is most likely that the NE feature is at the approaching side of the source. 

The arm-length ratio of the approaching to the receding lobes can be used to estimate the inclination angle ($i$) of the source, assuming there is no significant difference in the medium surrounding the jets on the two sides of the host galaxy. Using the equation of \cite{TaylorVermeulen}, the arm-length ratio can be given as
\begin{equation}
\frac{d_\mathrm{NE}}{d_\mathrm{SW}}=\frac{1+\beta\cos{i}}{1-\beta\cos{i}}
\end{equation}
From the VLA data of J1030$+$5516, the arm-length ratio is $1.81$, thus $\beta\cos{i}=0.29$, which gives a lower limit for the jet speed $\beta>0.3$, and an upper limit for the inclination angle $i<73\degr$. Similar equation describes the flux density ratio of the jet and counter-jet side \citep{TaylorVermeulen}:
\begin{equation}
\frac{S_\mathrm{NE}}{S_\mathrm{SW}}=\left(\frac{1+\beta\cos{i}}{1-\beta\cos{i}}\right)^{k-\alpha},
\end{equation}
where $k$ equals $2$ for continuous jet, and $3$ for discrete jet components. \cite{disc} using the FIRST data obtained a flux density ratio of $4.35$, from which using the spectral index $\alpha=-0.65$ they derived $\beta>0.2$, and  $i<79\degr$, which agree with our values obtained from the arm-length ratio. The higher resolution 5-GHz VLA data give a flux density ratio of the approaching and receding lobe of $2.85 \pm 1.8$, which agrees within the errors with the value \cite{disc} used. (Since we have no information on the spectral index of these radio features, we did not use it further to calculate the $\beta\cos{i}$ value.)

The Doppler factor is defined as,
\begin{equation}
\delta=\frac{1}{\gamma(1-\beta\cos{i})},
\end{equation}
where $\gamma=1/\sqrt{1-\beta^2}$ is the Lorentz factor.
Using the value $\beta\cos{i}=0.29$, and the lower limit on $\beta$, $0.29$, one can obtain an upper limit on the Doppler factor, $\delta<1.34$. 

\cite{doi-kpc} estimated the Doppler factor of a few radio-loud NLS1 sources by comparing the observed and intrinsic core powers. Fitting the VLA data we obtained the core flux density at $5$\,GHz. Assuming the spectral index of the core derived from the VLBA data ($\alpha_{4.3}^{7.6}=0.6$), the observed $5$-GHz power of the core is $3.6 \cdot 10^{24}$\,W\,Hz$^{-1}$. The intrinsic core power can be estimated using the empirical correlation found for radio galaxies, $\log P^\mathrm{core}_{5\mathrm{\,GHz}}=(0.62\pm 0.04) \log P^\mathrm{total}_{408\mathrm{\,MHz}}+ (7.6\pm 1.1)$, where $P^\mathrm{total}_{408\mathrm{\,MHz}}$ is the source's total power at $408$\,MHz \citep{gio_2001}. The closest frequency where the flux density of J1030$+$5516 was measured is $365$\,MHz within the framework of the Texas survey \citep{texas}. We used that value ($0.474\pm0.038$\,Jy) and the spectral index of the whole source derived by \cite{disc} to calculate $P^\mathrm{total}_{408\mathrm{\,MHz}}$. The corresponding intrinsic $5$-GHz core power is $\sim 10^{24}$\,W\,Hz$^{-1}$. Following \cite{doi-kpc}, if the difference between the observed and intrinsic core power is caused by relativistic beaming, their ratio can be given as $\delta^{3-\alpha}$. This implies a $\delta=1.7$. However, the uncertainty of the estimation of intrinsic core power allows for lower $\delta$ values and as high as $12$. \cite{disc} also used the above argument to estimate a Doppler factor. Instead of the core power, they used the intrinsic and observed core dominance parameter and obtained $\delta=3.3$. The derived $\delta$ values do not contradict the result of the VLBA observation which gave a lower limit of the brightness temperature. On the other hand, $\delta>1.34$ cannot be accommodated with the $\beta\cos{i}=0.29$ derived from the kpc-scale structure. Thus, if we accept the higher Doppler factors derived from the power of the core component, either the jet direction, $i$, or the jet speed, $\beta$, or both of them change significantly between the kpc and pc scale. The fact that only a single, compact radio-emitting feature was detected at mas-scale resolution indicates that the jet is not oriented close to the plane of the sky at pc scales.

The $1.4$-GHz radio power of the source calculated from the flux density detected in the FIRST survey ($\sim150$\,mJy) is $P_{\rm 1.4 GHz}^\mathrm{total} =9\cdot 10^{25}$\,W\,Hz$^{-1}$. \cite{AnBaan} studied the evolutionary sequence of symmetric extragalactic radio sources. On their radio power versus projected linear size diagram, J1030$+$5516 is among the large symmetric objects and the low-power FR\,II radio galaxies. 

The projected linear size of the source is $D\sim115$\,kpc. Assuming a constant expansion velocity and using the limit $\beta \cos i = 0.29$ derived from the kpc-scale radio structure, the kinematic age of the source can be estimated as 
\begin{equation}
t_\mathrm{kin}=\frac{D_{0}}{2\beta c}=\frac{D}{2 \beta c \sin i}=\frac{D}{2 \cdot 0.29 c \cdot \tan i}= \frac{6.3 \cdot 10^5}{\tan i}\mathrm{\,yr}
\end{equation}

where $D_0$ is the full (deprojected) size of the radio source. Using the upper limit on the inclination angle ($73\degr$), the lower limit on the age of the kpc-scale radio structure is $2\cdot 10^5$\,yr. The inclination angle of J1030$+$5516 should be below $\sim 32\degr$ to obtain an age $\gtrsim 10^6$\,yr and its age would reach $10^7$\,yr, if the inclination angle would be $\lesssim5\degr$, which is not consistent with a nearly symmetric kpc-scale structure. Thus, J1030$+$5516 seems to be younger than typical FRII radio galaxies whose lifetimes are estimated to be $(10^6-10^7)$\,yr \citep{Dea2009}.

\citet{Rakshit_WISE} investigated the infrared properties of $520$ NLS1 sources using {\it WISE} data. They found that more than $50$\% of the sources classified as variable in the AllWISE Source Catalog \citep{Cutri} fall within a specific region in the {\em WISE} color--color diagram, the `WISE Gamma-ray strip' (WGS). The WGS is defined by \citet{wgs1} and \citet{wgs2} as a distinct region where blazars are located. The infrared colors of J1030$+$5516 place this source within the WGS as well. Based upon this and its radio-loudness, \cite{dabrusco2014} included J1030$+$5516 in the list of $\gamma$-ray emitting blazar candidates. They defined three classes of blazar candidate sources based on their decreasing likeliness of being blazars. J1030$+$5516 fell into the second class.
This motivated us to analyse the available {\em Fermi}/LAT data to look for evidence of $\gamma$-ray emission. However, no $\gamma$-ray emission was found at the position of J1030$+$5516. 

\citet{disc} reported that J1030$+$5516 did not show variability in infrared at the wavelengths $3.4\mu$m and $4.6\mu$m measured with the {\it WISE} satellite. Since their publication, data from two additional epochs of {\em WISE} measurements became public. Our analysis of ten mission phases of the {\it WISE} data indicate a slight hint of long-term variability at the shorter wavelength due to the two additional epochs. There is no sign of short time scale (few day long) flux density changes neither in W1 nor in W2 bands.

Contrary to the finding of \citet{disc}, \citet{Graham2015} detected optical variability in J1030$+$5516. \citet{Graham2015} studied the optical light curves measured by the Catalina Real-time Transient Survey \citep{CRTS} to look for periodic variability in quasars, which they interpreted as induced by a closely separated binary supermassive black hole in those sources. They list J1030$+$5516 among the binary candidates with a period of $1515$ days, a separation of $0.006$\,pc and a rest-frame merger time of $2.2\cdot 10^5$\,yr (assuming a mass ratio of $0.5$). However, \citet{false_per} called for cautious approach when only a few cycles are used to assess periodic variability. In any case, whether periodic or not, J1030$+$5516 seems to show some variability in optical and infrared bands on time scales of years, as suggested by \citet{Graham2015} and perhaps also by our light curve compiled from {\em WISE} data (Fig.~\ref{fig:wise}).

\section{Summary and conclusion}

To reveal its structural properties at high resolution, we analysed archival radio interferometric observations of J1030$+$5516, a recently discovered rare radio-loud NLS1 source \citep{disc} having kpc-scale extended structure reminiscent of FR\,II radio galaxies. The compact central core and the nearly symmetric hotspots on its two sides seen in our 5-GHz VLA A-configuration image (Fig.~\ref{fig:VLA}) are consistent with the lower-resolution 1.4-GHz VLA FIRST survey image. A large fraction, more than a half of the radio emission originates from diffuse structures related to the two lobes, and is resolved out with the VLA at $5$\,GHz. At even higher, mas-scale resolution, the $4.6$ and $7.6$\,GHz VLBA observations show a single, unresolved core feature, with a brightness temperature $T_\mathrm{B}>3 \cdot 10^9$\,K. The observed high-resolution radio morphology and the derived parameters of J1030$+$5516 make it similar to the few known NLS1 sources with large-scale extended radio emission \citep{Richards_kpc, J1100}. 

Using the arm-length ratio, we derived a lower limit for the kpc-scale jet speed ($\beta>0.3$), and a corresponding upper limit of the inclination angle, $i<73\degr$, which are slightly more stringent than the ones given by \cite{disc}. The lack of complex radio morphology at mas scale indicates that the inner jet of J1030$+$5516 is not oriented close to the plane of the sky at pc scales.

The {\em WISE} infrared colors of the source place it in the WGS. However, our analysis of all available {\em Fermi}/LAT data did not reveal any $\gamma$-ray emission at the position of J1030$+$5516. There is indication of at least some variability on a multi-year time scale, both in the mid-infrared based on the most complete {\em WISE} light curve (Fig.~\ref{fig:wise}), and in the optical according to monitoring observations analysed by \citet{Graham2015}. Based on two epochs of radio observations, we also found indication of flux density variability.

Our results support the conclusion of \citet{disc} that the NLS1 galaxy J1030$+$5516 is likely a low-power and young version of an FR\,II radio galaxy with a double-lobed structure. There is no indication of relativistic boosting in J1030$+$5516, neither from radio imaging, nor from variability or $\gamma$-ray data, however it cannot be completely ruled out either.

\acknowledgments

This project received support from the Hungarian Research, Development and Innovation Office (OTKA NN110333). K.\'E.G. was supported by the J\'anos Bolyai Research Scholarship of the Hungarian Academy of Sciences. P.V. acknowledges support from {\it Fermi} grant NNM11AA01A. We used in our work VLBA data from project BP192F7 provided by the National Radio Astronomy Observatory that is a facility of the National Science Foundation operated under cooperative agreement by Associated Universities, Inc. We thank Leonid Petrov for providing results of data analysis prior to publication.
This publication makes use of data products from the Wide-field Infrared Survey Explorer, which is a joint project of the University of California, Los Angeles, and the Jet Propulsion Laboratory/California Institute of Technology, funded by the National Aeronautics and Space Administration.

\section*{}
Conflict of Interest: The authors declare that they have no conflict of interest.

\bibliographystyle{spr-mp-nameyear-cnd}

\end{document}